\documentclass[12pt]{article}

\newcommand{\cov}{{\rm {cov}}}

\newcommand{\br}{{\bf R}}
\newcommand{\Tr}{{{\rm Tr}}}

\usepackage{amsmath,amssymb,amsthm,amssymb}

\title{Parallelism  of quantum computations  from prequantum classical statistical field theory (PCSFT)}
\author{Andrei Khrennikov\\
School of Mathematics and Systems Engineering\\
University of V\"axj\"o, S-35195, Sweden}

\begin{document}

\maketitle

\abstract{This paper is devoted to such a fundamental problem of
quantum computing as quantum parallelism. It is well known that
quantum parallelism is the basis of the ability of quantum
computer to perform in polynomial time computations performed by
classical computers for exponential time. Therefore better
understanding of quantum parallelism is important both for
theoretical and applied research, cf. e.g. David Deutsch
\cite{DD}.  We present a realistic interpretation based on
recently developed prequantum classical statistical field theory
(PCSFT). In the PCSFT-approach to QM quantum states (mixed as well
as pure) are labels of special ensembles of classical fields. Thus
e.g. a single (!) ``electron in the pure state'' $\psi$ can be
identified with a special `` electron random field,'' say
$\Phi_\psi(\phi).$ Quantum computer operates with such random
fields. By one computational step for e.g. a Boolean function
$f(x_1,...,x_n)$ the initial random field $\Phi_{\psi_0}(\phi)$ is
transformed into the final random field $\Phi_{\psi_f}(\phi)$
``containing all values'' of $f.$ This is the objective of quantum
computer's ability to operate quickly with huge amounts of
information -- in fact, with classical random fields.}


\section{Introduction}
Recent tremendous development of quantum information theory and
especially quantum computing and cryptography stimulated research
in foundations of quantum mechanics, see, e.g.,  \cite{ADC1} -- \cite{P3},
problems which have been of merely theoretical (or even
philosophic) interest became extremely important for understanding
of processing of quantum information. Nowadays quantum foundations
 have important implications for engineering and nanotechnology. In
this note we discuss one exciting problem of quantum computing,
namely, {\it quantum parallelism.} It is well known that quantum
algorithms could solve in polynomial time some problems which need
exponential time for known classical algorithms. The ability to
reduce essentially computational time is one of the main
motivations for development of quantum computers. This ability is
closely related to one of the main problems in quantum
foundations, namely interpretation of {\it superposition of
quantum states.} In quantum computing the ability to operate with
superpositions is called quantum parallelism.

We emphasize that
any realistic interpretation of
quantum parallelism which is accepted by the majority of the quantum information community
has not yet been provided. Of course, one may just follow the orthodox
Copenhagen interpretation. But it is not a realistic one.
Some creators of quantum computing were not satisfied by the
orthodox Copenhagen interpretation of quantum parallelism. An
interesting attempt to provide a kind of realistic understanding
of quantum parallelism was done by Deutsch\cite{DD}  who used
the {\it many worlds} interpretation of quantum computing. In the
many worlds approach quantum parallelism can be understood in the
realistic way as classical parallelism in many worlds. However,  the
many worlds interpretation is not (at leat yet) commonly  accepted. The quantum
computing majority would prefer the orthodox Copenhagen
interpretation.

Recently I proposed a new realistic ground for QM, see \cite{KL34}-\cite{KHY4}. It
was shown that QM can be represented as an asymptotic projection
of classical statistical mechanics with {\it infinite dimensional
phase space.} By representing this space as
$$\Omega=L_2 (\br^3) \times L_2 (\br^3)$$ we represent its points
by classical vector fields $\phi(x)=(q(x), p(x)).$  In our model -
prequantum (classical statistical) field theory (PCSFT) - these
classical field might be considered as a {\it kind} of hidden
variables. The corresponding complex representation of such vector
fields is given Riemann-Silberstein vector (which is used for the
complex representation of the classical electrodynamics; in
particular, Maxwell equations in  empty space are transformed into
Schr\"odinger's type equation):
$$
\phi(x)= q(x) +i p(x).
$$

The correspondence rules between PCSFT and QM differ from rules
which are typically considered in theories with hidden variables,
e.g., rules which are formalized in known ``NO-GO''  theorems (von
Neumann, Kochen-Specker, Bell,...). A classical field $\phi(x)$ of
PCSFT does not determine the values of conventional quantum
observables (as it should be in a theory with hidden variables).

In the PCSFT-framework we define so called {\it prequantum variables}
which are given by functionals $f(\phi)$ of classical fields.
Corresponding (conventional) quantum observable $\hat A$ is given
by the second derivative of a functional (``classical variable'')
$f(\phi).$ We set  $\hat A\equiv T(f) = f^{\prime\prime}(0)/2.$ On the one hand,
 this rule for correspondence between prequantum  variables and quantum observables,
 $f \to T(f),$ satisfies an important assumption of the von Neumann ``NO-GO''  theorem:
$$
T(f_1+...+f_n)=  \hat A_1 +...+ \hat A_n,
$$
where operators $ \hat A_1, ..., \hat A_n$ need not commute, \cite{VN}. On the other hand, the
``spectral postulate'' is violated: the ranges of values of a prequantum variable
$f(\phi)$ and the corresponding quantum observable $\hat A\equiv T(f)$ do not coincide.

Nevertheless,  average with respect to a classical random field  $\Phi(\phi)$ (here $\phi \in \Omega$ plays the role of
the random parameter):
$$<f>_\Phi\equiv E f(\Phi(\phi))$$  can be
approximated by quantum average given by the von Neumann
trace-formula, \cite{VN}: $$<A>_ \rho\equiv \Tr \;\rho \; \hat
A.$$   Here the operator $\rho$ is obtained by normalization of
the covariance operator of the random field $\Phi(\phi).$ It has
all features of the von Neumann density operator, \cite{VN}. The
quantum average $<A>_ \rho$ gives the first order approximation
of classical field average $<f>_\Phi.$ The small parameter of this
asymptotic expansion of $<f>_\Phi$ is given by dispersion of the
random field:
$$
\kappa= E \Vert \Phi(\phi) \Vert^2,
$$
where $\Vert \cdot \Vert$ is the norm on the Hilbert space $\Omega.$

To distinguish PCSFT from conventional theories with hidden
variables, we shall call classical fields $\phi \in \Omega$  {\it
ontic hidden variables,} cf.  \cite{ATM}.  ``Ontic'' does not mean
that effects such variables could not be measured in principle.
However, to find effects of ``prequantum fields'' one should
develop new measurement technologies corresponding to ``prequantum
variables'' given by functionals of classical fields (electron
field, proton field and so on). So, for such more advanced
technologies they will become observables.

We point out that a similar problem is present in some other
models with hidden variables, but typically not so much attention is paid
 to it. For example, in one of the most known models of
this type -- Bohmian mechanics -- momentum is not the conventional
momentum of QM. I would consider Bohmian mechanics as a model with
``semi-ontic'' hidden variables. Position $q$ is the conventional
hidden variable, but momentum $p$ is the ontic one.

In our approach quantum states (mixed as well as pure) are images
of ensembles of classical fields. They are  mathematically
described by measures $\mu$ on $\Omega$. Such ensembles can be
considered as {\it random fields,} see also \cite{DAV}. Therefore in PCSFT
quantum computations can be represented as processing of random
fields. Quantum parallelism is classical parallelism, but for
ensembles of fields.

It might be that we found the main source of tremendous ability of
quantum computer to operate with huge amounts of information. In
contrast to classical computer, quantum computer operates with
infinite dimensional objects composing infinitely large ensembles.

One may say that operations with quantum bits are  performed in
the finite dimensional Hilbert spaces. However, the same arguments
that I. V. Volovich presented in \cite{VL}  for quantum cryptography
should also be applied to quantum computing. Quantum computer
operates in physical space -- by our model with physical fields
which are infinite dimensional objects.

\subsection{Copenhagen interpretations}

The main distinguishing feature of the Copenhagen interpretation
is association of the wave function $\psi$ -- pure quantum state
-- with an {\it individual quantum system.} For example, one
(e.g., Heisenberg, Pauli, Dirac, Fock or Landau)\footnote{It is
not clear at all whether Bohr would assign him self  \cite{Bohr} to such an
interpretation, see  Plotnitsky \cite{P}--\cite{P3} for discussion.} would
speak about the wave function of the electron or in other words
about the electron having the concrete pure state $\psi.$

Opposite to such an individual interpretation, by the ensemble
interpretation (e.g., Einstein and Ballentine, \cite{BL1}, \cite{BL2}) the wave
function $\psi$ is associated not with an individual quantum
system, but with an {\it ensemble of quantum systems} prepared
under the same complex of experimental physical conditions --
preparation procedure, \cite{BL1}, \cite{BL2}, \cite{Holevo}.  By the ensemble interpretation
there is no difference between pure and mixed quantum states. Pure
quantum states represent ensembles of systems as well as mixed
states.

By the Copenhagen interpretation QM is complete. One could not
introduce a mathematical model in which  quantum systems are
described by hidden variables determining the values of quantum
observables. By the ensemble interpretation such variables can be
introduced. The wave function $\psi$ is just a mathematical symbol
for an ensemble of quantum systems and each system $s$ in this
ensemble can be characterized by a value of the hidden variable
$\lambda\equiv \lambda_s.$

Typically the Copenhagen interpretation is considered as a
``NO-GO''-interpretation. It seems that its completeness does not
permit a more detailed description of physical reality
than the one given by QM. On the other hand, the ensemble
interpretation does not claim that QM is complete and that the QM
description is the final one.

\medskip

{\it Is the Copenhagen interpretation really a
``NO-GO''-interpretation?}

\medskip

It is correct, but only to some extend. In fact, we can proceed in
the following way.

Let us interpret quantum particles as {\it classical random
fields.} For example, any electron is a classical random
field\footnote{Thus in such a model we have a variety of fields
corresponding to different ``quantum particles'',  e.g., the
electron field  or the proton field.} $\Phi(x,\omega),$ where
$x\in {\bf R}^3$ and $\omega$ is a random parameter. In
our approach the wave function $\psi$
(describing by the Copenhagen interpretation a pure quantum state)  is considered
 as some characteristic of the corresponding field
$\Phi(x,\omega).$ As usual in random field theories, one can choose the
random parameter $\omega=\phi \in \Omega.$ We can not introduce a conventional hidden variable
associated with the electron and determining values of quantum
observables. The concrete value is the result of interaction of
the random field with the corresponding measuring device. An
individual fluctuation $\Phi(x,\omega_0)$ (i.e., for the fixed
$\omega_0)$ does not determine this value. As was pointed out in
introduction it is more natural to call prequantum fields ontic
hidden variables.

We remark that at each preparation act a preparation procedure
produces not a single fluctuation of e.g. the electron field, but
an ensemble of electron fields. Such an ensemble has the label
``electron'' in QM.

\medskip

This interpretation we call the {\it random field Copenhagen
interpretation.}

\medskip

The simplest model of such type can be obtained by considering two
time scales , \cite{QT}:

a) a quick time scale (``prequantum time scale'');

b) slow time scale -- the scale of measurements (``quantum time
scale'')\footnote{By following Bohr \cite{Bohr}  I consider QM as theory of
measurements,  see e.g. Plotnitsky \cite{P}--\cite{P3} for details.  Therefore it is
natural for me to call the time scale of measurements
the quantum time scale.  I  understood well  that this terminology might be misleading,
since many authors (especially in quantum cosmology and string theory)  use the terminology
the quantum time scale for  the Planck time scale.}

Denote quick and slow times by symbols $s$ and $t,$  respectively.
A random field describes fluctuations on the prequantum time
scale. Such fluctuations are not visible at the quantum time
scale. Measurements are averages with respect to $s$-time. An
instant of quantum (laboratory, physical) time correspond to a
huge interval of prequantum time.

In this model it is clear why prequantum observables are different
from  quantum ones. These are two classes of observables
corresponding to two different time scales.

\medskip

The main problem for justification of the random field Copenhagen
interpretation is to create  a random field model which would
couple in a natural way classical random field averages with
averages given by the QM-formalism, namely, by von Neumann's trace
formula:
$$
<\hat A>_\rho= \rm{Tr} \; \rho \hat A.
$$

\section{Prequantum classical statistical field theory}

We define {\it ``classical statistical models''} in the following
way: a) physical states $\omega$ are represented by points of some
set $\Omega$ (state space); b) physical variables are represented
by functions $f: \Omega \to {\bf R}$ belonging to some functional
space $V(\Omega);$ c) statistical states are represented by
probability measures on $\Omega$ belonging to some class
$S(\Omega);$ d) the average of a physical variable (which is
represented by a function $f \in V(\Omega))$ with respect to a
statistical state (which is represented by a probability measure
$\rho \in S(\Omega))$ is given by
\begin{equation}
\label{AV0} < f >_\rho \equiv \int_\Omega f(\phi) d \rho(\phi) .
\end{equation}
A {\it classical statistical model} is a pair $M=(S,
V).$\footnote{ We recall that classical statistical mechanics on
the phase space $\Omega_{2n}= {\bf R}^n\times {\bf R}^n$ gives an
example of a classical statistical model. But we shall not be
interested in this example in our further considerations. We shall
develop  a classical statistical model with {\it an
infinite-dimensional phase-space.}}

The conventional quantum statistical model with the complex
Hilbert state space $\Omega_c$ is described in the following way:
a) physical observables are represented by operators $A: \Omega_c
\to \Omega_c$ belonging to the class of continuous    self-adjoint
operators\footnote{Of course, discontinous (unbounded) operators
are important in QM. However, as was pointed by von Neumann
\cite{VN}, it is always possible to restrict consideration to
continuous operators, since discontinuous ones can be approximated
by continuous ones.} ${\cal L}_s \equiv {\cal L}_s (\Omega_c);$ b)
statistical states are represented by von Neumann density
operators (the class of such operators is denoted by ${\cal D}
\equiv {\cal D} (\Omega_c));$ d) the average of a physical
observable (which is represented by the operator $A \in {\cal L}_s
(\Omega_c))$ with respect to a statistical state (which is
represented
  by the density operator $D \in {\cal D} (\Omega_c))$ is given by von Neumann's
formula:
\begin{equation}
\label{AV1} <A >_D \equiv \rm{Tr}\; DA
\end{equation}
The {\it quantum statistical model} is the pair $N_{\rm{quant}}
=({\cal D}, {\cal L}_s).$

\medskip

We are looking for a classical statistical model $M=(S, V)$ which
will provide {\it ``dequantization'' of the quantum model}
$N_{\rm{quant}} =({\cal D}, {\cal L}_s).$ By dequantization we
understand constructing of a classical statistical model such that
averages given by this model can be approximated by quantum
averages. Approximation is based on the asymptotic expansion of
classical averages with respect to a small parameter. The main
term of this expansion coincides with the corresponding quantum
average.

We choose the phase space $\Omega= Q\times P,$ where $Q=P=H$ and
$H$ is the real (separable) Hilbert space.  We consider $\Omega$ as
the real Hilbert space with the scalar product $(\phi_1, \phi_2)=
(q_1, q_2) + (p_1, p_2).$ We denote  by $J$ the symplectic
operator on $\Omega:
 J= \left( \begin{array}{ll}
 0&1\\
 -1&0
 \end{array}
 \right ).$
Let us consider the class ${\cal L}_{\rm symp} (\Omega)$ of
bounded ${\bf R}$-linear operators $A: \Omega \to \Omega$ which
commute with the symplectic operator:\begin{equation}
\label{SS}
A J= J A .
\end{equation}
This is a subalgebra of the algebra of bounded linear operators
${\cal L} (\Omega).$ We also consider the space of ${\cal
L}_{\rm{symp}, s}(\Omega)$ consisting of self-adjoint operators.

By using the operator $J$ we can introduce on the phase space
$\Omega$ the complex structure. Here $J$ is realized as $-i.$ We
denote $\Omega$ endowed with this complex structure by $\Omega_c:
\Omega_c\equiv Q\oplus i P.$ We shall use it later. At the moment
consider $\Omega$ as a real linear space and consider its
complexification $\Omega^{{\bf C}}= \Omega \oplus i \Omega.$

Let us consider the functional space ${\cal
V}_{\rm{symp}}(\Omega)$ consisting of functions $f:\Omega \to {\bf
R}$ such that: a) the state of vacuum is preserved\footnote{The
vacuum state is such a classical field which amplitude is zero at
any point $x.$} : $f(0)=0;$ b) $f$ is $J$-invariant: $f(J\Phi)=
f(\phi);$ c) $f$ can be extended to the analytic function
$f:\Omega^{{\bf C}}\to {\bf C}$ having the exponential growth: $
\vert f(\phi)\vert \leq c_f e^{r_f \Vert \Phi \Vert} $ for some
$c_f, r_f \geq 0$ and for all $\phi\in \Omega^{{\bf C}}.$

The following trivial mathematical result plays the fundamental
role in establishing classical $\to$ quantum correspondence: {\it
Let $f$ be a smooth $J$-invariant function. Then } $f^{\prime
\prime}(0)\in {\cal L}_{\rm{symp}, s}(\Omega).$ In particular, a
quadratic form is $J$-invariant iff it is determined by an
operator belonging to ${\cal L}_{\rm{symp}, s}(\Omega).$

We consider the space statistical states $S_{G,
\rm{symp}}^{\kappa}(\Omega)$ consisting of measures $\rho$ on
$\Omega$ such that: a) $\rho$ has zero mean value; b) it is a
Gaussian measure; c) it is $J$-invariant; d) its dispersion has
the magnitude $\kappa.$ Thus these are $J$-invariant Gaussian
measures such that $$ \int_\Omega \Phi d\rho(\phi)=0 \;
\mbox{and}\; \sigma^2(\rho)= \int_\Omega \Vert \Phi\Vert^2 d
\rho(\phi)= \kappa, \; \kappa \to 0.
$$
Such measures describe small Gaussian fluctuations. The following
trivial mathematical result plays the fundamental role in
establishing classical $\to$ quantum correspondence: {\it Let a
measure $\rho$ be $J$-invariant. Then its covariation operator}
$B= \rm{cov}\; \rho \in {\cal L}_{\rm{symp}, s}(\Omega).$ Here
$$(By_1, y_2)= \int (y_1, \Phi)(y_2, \Phi) d \rho( \Phi).$$

We now consider the complex realization $\Omega_c$ of the phase
space and the corresponding complex scalar product $<\cdot,
\cdot>.$ We remark that the class of operators ${\cal L}_{\rm
symp} (\Omega)$ is mapped onto the class of ${\bf C}$-linear
operators ${\cal L}(\Omega_c).$ We also remark that, for any $A\in
{\cal L}_{\rm{symp}, s}(\Omega),$ real and complex quadratic forms
coincide:
$ (A\Phi,\phi) =<A\Phi,\phi>.$
We also define for any measure its complex covariation operator
$B^c= \rm{cov}^c \rho$ by $$ <B^c y_1, y_2>=\int <y_1, \Phi>
<\phi, y_2> d \rho (\phi).$$ We remark that for a $J$-invariant
measure $\rho$ its complex and real covariation operators are
related as $B^c=2 B.$ As a consequence, we obtain that any
$J$-invariant Gaussian measure is uniquely determined by its
complex covariation operator. As in the real case [1], we  can
prove that for any operator $ A\in {\cal L}_{\rm{symp},
s}(\Omega):$
$\int_\Omega <A\Phi,\phi> d \rho (\phi) = \rm{Tr} \;\rm{cov}^c
\rho \;A.$
 We point out that the trace is considered with respect to the complex
inner product.

 We consider now the one parameter family of classical statistical
models:
\begin{equation}
\label{MH} M^\kappa= ( S_{G, \rm{symp}}^\kappa(\Omega),{\cal
V}_{\rm{symp}}(\Omega)), \; \kappa\geq 0,
\end{equation}

By making in the Gaussian infinite-dimensional integral the change
of variables (field scaling):
\begin{equation}
\label{ANN3Z} \Phi= \sqrt{\kappa} \Psi,
\end{equation}
we obtain the following result \cite{KHY4}:

\medskip

{\it Let $f \in {\cal V}_{\rm{symp}}(\Omega)$ and let $\rho \in
S_{G, \rm{symp}}^\kappa(\Omega).$ Then the following asymptotic
equality holds:
\begin{equation}
\label{ANN3} <f>_\rho =  \frac{\kappa}{2} \; \rm{Tr}\; D^c \;
f^{\prime \prime}(0) + O(\kappa^2), \; \kappa \to 0,
\end{equation}
where the operator $D^c= \rm{cov}^c \; \rho/\kappa.$ Here
\begin{equation}
\label{OL} O(\kappa^2) = \kappa^2 R(\kappa, f, \rho),
\end{equation}
where $\vert R(\kappa,f,\rho)\vert \leq c_f\int_\Omega  e^{r_f
\Vert \Psi \Vert}d\rho_{D^c} (\Psi).$ }

\medskip
Here $\rho_{D^c}$ is the Gaussian measure with zero mean value and
the complex covariation operator $D^c.$

We see that the classical average (computed in the model
$M^\kappa= ( S_{G, \rm{symp}}^\kappa(\Omega),{\cal
V}_{\rm{symp}}(\Omega))$ by using the measure-theoretic approach)
is coupled through (\ref{ANN3}) to the quantum average (computed
in the model $N_{\rm{quant}} =({\cal D}(\Omega_c),$ ${\cal
L}_{{\rm s}}(\Omega_c))$ by the von Neumann trace-formula).

The equality (\ref{ANN3}) can be used as the motivation for
defining the following classical $\to$ quantum map $T$ from the
classical statistical model $M^\kappa= ( S_{G,
\rm{symp}}^\kappa,{\cal V}_{\rm{symp}})$ onto the quantum
statistical model $N_{\rm{quant}}=({\cal D}, {\cal L}_{{\rm s}}):$
\begin{equation}
\label{Q20} T: S_{G, \rm{symp}}^\kappa(\Omega) \to {\cal
D}(\Omega_c), \; \; D^c=T(\rho)=\frac{\rm{cov}^c \; \rho}{\kappa}
\end{equation}
(the Gaussian measure $\rho$ is represented by the density matrix
$D^c$ which is equal to the complex covariation operator of this
measure normalized by  $\kappa$);
\begin{equation}
\label{Q30} T: {\cal V}_{\rm{symp}}(\Omega) \to {\cal L}_{{\rm
s}}(\Omega_c), \; \; A_{\rm quant}= T(f)= \frac{1}{2}
f^{\prime\prime}(0).
\end{equation}
Our previous considerations can be presented in the following form
\cite{KHY4}

\medskip

{\bf Beyond QM Theorem.}  {\it The one parametric family of
classical statistical models $M^\kappa= ( S_{G,
\rm{symp}}^\kappa(\Omega),{\cal V}_{\rm{symp}}(\Omega))$ provides
dequantization of the quantum model $N_{\rm{quant}} =({\cal
D}(\Omega_c),$ ${\cal L}_{{\rm s}}(\Omega_c))$ through the pair of
maps (\ref{Q20}) and (\ref{Q30}). The classical and quantum
averages are coupled by the asymptotic equality (\ref{ANN3}).}

\section{The random field Copenhagen
interpretation for PCSFT}

In the series of papers  \cite{KL34}-\cite{KHY4} I used the  ensemble interpretation
(in the spirit of Einstein, Margenau, Ballentine) to couple my
model with ontic  hidden variables PCSFT with QM. By this interpretation
a classical statistical state $\mu$ of a prequantum theory
represents an ensemble of  hidden variables,  say $E_\mu.$ In our
case the theory with (ontic) hidden variables is PCSFT and hidden
variables are classical fields, $\phi(x), x \in {\bf R}^3.$ In
\cite{KL34}-\cite{KHY4}  I did not proceed carefully and I did not distinguish
conventional hidden variables form the ontic ones, see the discussion
in introduction.

If a quantum system, e.g., an electron, has the quantum state $
\rho =T(\mu)=\cov^c \mu,$ then we assumed that, in fact, his state
is given by the fixed field $\phi \in E_\mu.$ By considering an
ensemble of electrons prepared in the state $\rho$ we reproduce
the ensemble of classical fields $E_\mu.$

Recently I found that such an ensemble interpretation is not the
only possible interpretation for coupling of PCSFT with QM.
Surprisingly PCSFT-QM coupling could also be interpreted in the
Copenhagen's way.

We recall that by the orthodox Copenhagen interpretation a pure
quantum state $\psi, \Vert \psi\Vert =1,$ describes an {\it
individual quantum system.}

This interpretation of QM we combine with PCSFT in the  following
way. Suppose that $\mu \equiv \mu_\psi$ has the covariance
operator $$B=\kappa \psi \otimes \psi.$$ In the new interpretation
$\mu_\psi$ represents a random field: "mixture of fields $\phi \in
\Omega$ with weights $\mu_\psi(\phi).$"

We emphasize that there is considered mixture and not
superposition. We denote this random field by $\Phi_\psi(\phi);$
here $\phi \in \Omega$ plays the role of a random parameter.

We take a quantum system having the QM-state $\psi,$ e.g., an
electron in the state $\psi.$ We  now consider $\psi$ as the
symbol denoting the random field $\Phi_\psi(\phi).$ This is
nothing else than {\it random field Copenhagen interpretation.} By
such an interpretation each quantum system is nothing else than a
mixture of classical fields.\footnote{There is no problem with the superposition principle.
It holds true for any classical prequantum field.}
 At the moment (e.g. due to
technological problems)  we are not able to distinguish those
fields. By identifying a quantum system with a random field we
explain the origin of quantum randomness. Opposite to von
Neumann \cite{VN}, we do not consider quantum randomness as irreducible.

We remark that if one uses the {\it random field Copenhagen
interpretation} then quantum randomness is not reduced to
randomness for an ensemble of quantum systems, e.g., electrons.
Nevertheless, in our approach quantum randomness is reduced to
ensemble randomness, namely, to ensembles of classical fields.

\section{Quantum parallelism}
The main distinguishing feature  of quantum computation, see,
e.g., Simon's algorithm, is the possibility to prepare the
quantum state

\begin{equation}
\label{X} \psi_0 = \frac{1}{2^{n/2}} \sum_x |x>,
\end{equation}
containing all possible values of the argument $x,$ and then to
transform the state $\psi_0$ into the quantum state
\begin{equation}
\label{X1} \psi_f=\sum_x|x>|f(x)>,
\end{equation}
containing all  values of $f.$ The possibility to create the state
$\psi_f$ by one step of quantum computations (by using oracal
$U_f$) implies the possibility to perform on quantum computer in
polynomial time calculations which are done in nonpolynomial time
on classical computer.

Typically the difference between {\it quantum parallelism and
classical parallelism} is emphasized. For example, in the book of
A. S. Holevo \cite{HV}, he pointed out that all values $f(x)$ are present
in $\psi_f$ in the latent form and one should not identify this
latent presence with the result of parallel computations on a
classical computer. (In the latter case there are really produced
all values $f(x).$)

In PCSFT the gap between quantum parallelism and classical
parallelism is essentially less. By the random field Copenhagen
interpretation states $\psi_0$ and $\psi_f$ are symbols denoting
random fields $\Phi_{\psi_0}$ and $\Phi_{\psi_f}.$  Therefore all
values of the argument $x$ are really present in the ensemble of
classical fields $\Phi_{\psi_0}(\phi), \phi \in \Omega.$ Oracle
$U_f$ really transfers these values into corresponding values of
$f$ which are all contained in the random field $\Phi_{\psi_f}
(\phi), \phi \in \Omega.$

We remark that by PCSFT, see \cite{KL34}--\cite{KHY4},
$$
\Phi_{\psi}(U_t\Phi)=\Phi_{U_t\psi}(\phi),
$$
for any one parametric group of unitary operators $U_t$ and hence, in particular,
for any unitary operator $U_f$ representing quantum computation.

The crucial difference from the classical parallelism is that we
are not able to extract all these values from the final random
field. A measurement destroys the structure of a random field.
Therefore, to repeat this measurement, we should produce a new
random field.

\medskip

{\bf Conclusion.} {\it Quantum  parallelism can be interpreted in
the realistic way in the framework of PCSFT. By the random field
Copenhagen  interpretation this is parallelism of computations
over ensembles of classical fields - random fields.}

Some results of this paper  were presented at the seminar of the
Quantum Field Section of Steklov Mathematical Institute and at the
seminar on Quantum Computers of Institute of Physics and
Technology of Russian Academy of Science. The
author is grateful to all participants of these seminars for
fruitful discussions and especially for comments, critical remarks
and advices given by A.  Slavnov, B. Dragovich, K. Valiev, A.
 Holevo, Yu.  Bogdanov and I. Basieva.


\begin{thebibliography}{99}

\bibitem{DD} Deutsch,  D.,  {\it The fabric of reality,} The penguin Press,
Alllen Lane

\bibitem{ADC1} Adenier, G.,  Khrennikov, A. Yu. (eds): Foundations of Probability and
Physics-3.  American Institute of Physics, Ser. Conference
Proceedings 750, Melville, NY (2005)

\bibitem{ADC2} Adenier, G.,  Khrennikov, A. Yu. and Nieuwenhuizen, Th.M. (eds.):
Quantum Theory: Reconsideration of Foundations-3. American
Institute of Physics, Ser. Conference Proceedings 810, Melville,
NY (2006)


\bibitem{ADC3} Adenier, G., Fuchs, C. and Khrennikov, A. Yu. (eds): Foundations of Probability and
Physics-3.  American Institute of Physics, Ser. Conference
Proceedings 889, Melville, NY (2007)

\bibitem{AA} Aaronson, S.: Is quantum mechanics an island in
theoryspace? In: Khrennikov, A.Yu. (ed) Quantum Theory:
Reconsideration of Foundations -2, pp. 15-28. Ser. Math. Model.
10, V\"axj\"o University Press, V\"axj\"o (2003); electronic
volume: http://www.vxu.se/msi/forskn/publications.html

\bibitem{ATM} Atmanspacher, H. and Primas, H.: Epistemic and ontic quantum
realities. In:  Adenier, G.,  Khrennikov, A. Yu. (eds) Foundations
of Probability and Physics-3, pp. 49-62.  American Institute of
Physics, Ser. Conference Proceedings 750, Melville, NY (2005)

\bibitem{Bacci} Bacciagaluppi,  G.: Classical extensions, classical representations and Bayesian
updating in quantum mechanics.  In: Khrennikov, A.Yu. (ed) Quantum
Theory: Reconsideration of Foundations -2, pp. 55-70. Ser. Math.
Model. 10, V\"axj\"o University Press, V\"axj\"o (2003);
electronic volume: http://www.vxu.se/msi/forskn/publications.html

\bibitem{BL3} Ballentine, L. E.:  Interpretations of probability and
quantum theory. In:  Khrennikov, A. Yu. (ed)  Foundations of
Probability and Physics, Quantum Probability and White Noise
Analysis,  13, pp. 71-84. WSP, Singapore (2001)

\bibitem{BIAL} Bialynicki-Birula, I.: R\'enyi entropy and the
uncertainty relations. In: Adenier, G., Fuchs, C. and Khrennikov,
A. Yu. (eds) Foundations of Probability and Physics-3, pp. 52--61.
American Institute of Physics, Ser. Conference Proceedings 889,
Melville, NY (2007)

\bibitem{BUS1} Busch, P.: Less (precision) is more (information):
Quantum information in fuzzy probability theory. In: Khrennikov,
A.Yu. (ed) Quantum Theory: Reconsideration of Foundations -2, pp.
113-148. Ser. Math. Model. 10, V\"axj\"o University Press,
V\"axj\"o (2003); electronic volume:
http://www.vxu.se/msi/forskn/publications.html

\bibitem{DARIANO} D´Ariano, G.M.: Operational axioms for quantum mechanics. In: Adenier, G., Fuchs, C. and Khrennikov, A. Yu. (eds)
Foundations of Probability and Physics-3, pp. 79--105. American
Institute of Physics, Ser. Conference Proceedings 889, Melville,
NY (2007)

\bibitem{Folse} Folse, H.J.: Bohr's conception of the quantum mechanical state of a system and its role in the framework
of complementarity. In: Khrennikov, A.Yu. (ed) Quantum Theory:
Reconsideration of Foundations, , pp. 83-98. Ser. Math. Model. 2,
V\"axj\"o University Press, V\"axj\"o  (2002); electronic volume:
http://www.vxu.se/msi/forskn/publications.html


\bibitem{ME1} Mermin, N.D.: Whose knowledge?
In: Khrennikov, A.Yu. (ed) Quantum Theory: Reconsideration of
Foundations. Ser. Math. Model. 2, pp. 261-270, V\"axj\"o
University Press, V\"axj\"o  (2002); electronic volume:
http://www.vxu.se/msi/forskn/publications.html


\bibitem{PERREZ} Perez-Suarez, M., Santos, D.J.; Quantum mechanics
as an information theory: Some further missing on a Fuchsian
proposal. In: Khrennikov, A.Yu. (ed) Quantum Theory:
Reconsideration of Foundations, pp. 469-478. Ser. Math. Model. 2,
V\"axj\"o University Press, V\"axj\"o  (2002); electronic volume:
http://www.vxu.se/msi/forskn/publications.html

\bibitem{P} Plotnitsky, A.:  Reading Bohr: Complementarity, Epistemology, Entanglement,
and Decoherence. In: Gonis, A. and Turchi, P. (eds) NATO Workshop
Decoherence and its Implications for Quantum Computations,
pp.3--37.   IOS Press, Amsterdam (2001)

\bibitem{P1} Plotnitsky, A.: The Knowable and Unknowable: Modern Science,
Nonclassical Thought, and the ``Two Cultures.'' Univ. Michigan
Press (2002)

\bibitem{P2} Plotnitsky, A.:  ``This is an extremely funny thing,
something must be hidden behind that'': Quantum waves and quantum
probability with Erwin Schr\"odinger. In: Adenier, G., Khrennikov,
A. Yu. (eds) Foundations of robability and physics---3, pp.
388--408. American Institute of Physics, Ser. Conference
Proceedings 750, Melville, NY (2005)

\bibitem{P3} Plotnitsky, A.: Reading Bohr: Physics and Philosophy. Springer, Dordrecht (2006)

\bibitem{KL34} Khrennikov, A.Yu.: A pre-quantum classical statistical model
with infinite-dimensional phase space. J. Phys. A: Math. Gen. 38,
9051-9073 (2005).

\bibitem{KL38} Khrennikov, A.Yu.: Generalizations of quantum mechanics induced by classical statistical field
theory. Found. Phys. Lett.  18, 637-650 (2006)

\bibitem{KL46} Khrennikov, A.Yu.: Nonlinear Schrödinger equations from prequantum classical statistical
field theory. Phys. Lett. A  357, 171-176 (2006)

\bibitem{KL47} Khrennikov, A.Yu.: Prequantum classical statistical field theory: Complex representation,
Hamilton-Schr\"odinger equation, and interpretation of stationary
states. Found. Phys. Lett. 19,  299-319 (2006)

\bibitem{KHY4} Khrennikov, A. Yu.:  Quantum mechanics as an asymptotic
projection of statistical mechanics of classical fields. In: Adenier, G., Khrennikov,
 A. Yu. and Nieuwenhuizen, Th.M. (eds) Quantum theory: reconsideration of foundations---3,  pp. 179--197.
American Institute of Physics, Ser. Conference Proceedings 810,
Melville, NY (2006).

\bibitem{VN}  Von Neuman, J.: Mathematical Foundations of
Quantum Mechanics. Princeton University Press, Princeton (1955)

\bibitem{DAV} Davidson, M.P.: Stochastic models of quantum
mechanics - a perspective.  In: Adenier, G., Fuchs, C. and
Khrennikov, A. Yu. (eds) Foundations of Probability and Physics-3,
pp. 106--119. American Institute of Physics, Ser. Conference
Proceedings 889, Melville, NY (2007)

\bibitem{VL} Volovich, I.V.: Quantum cryptography in space and Bell's theorem.
In: Khrennikov, A,Yu. (ed) Foundations of probability and physics,
pp. 364--372. QP--PQ: Quantum Prob. White Noise Anal. 13. WSP,
River Edge, NJ (2001)

\bibitem{Bohr}  N. Bohr, {\it The philosophical writings of Niels
Bohr}, 3 vols. (Woodbridge, Conn., Ox Bow Press, 1987)

\bibitem{BL1} Ballentine, L.E.: The statistical interpretation of quantum mechanics.
Rev. Mod. Phys. 42, 358--381 (1989)

\bibitem{BL2} Ballentine, L.E.: Quantum Mechanics. Prentice Hall, Englewood Cliffs, NJ (1989)

\bibitem{Holevo} Holevo, A. S.:  Statistical Structure of Quantum Theory.
Springer, Berlin-Heidelberg (2001)

\bibitem{QT} A. Khrennikov, Quantum Randomness as a Result of Random
Fluctuations at the Planck Time Scale? {\it Int. J. Theor. Phys.},
{\bf 47}, N.1, (2008).

\bibitem{HV} Holevo, A. S.: Introduction to Quantum Information Theory.
Moscow State Univ. Publ., Moscow, 2003.

\end{thebibliography}
\end{document}